\documentclass{article}
\usepackage{ltwol2e}
\usepackage{epsfig}

\def\Journal#1#2#3#4{{#1} {\bf #2}, #3 (#4)}

\def\NPB{{\em Nucl. Phys.} B}
\def\PLB{{\em Phys. Lett.}  B}
\def\PRL{\em Phys. Rev. Lett.}
\def\PRD{{\em Phys. Rev.} D}
\def\ZPC{{\em Z. Phys.} C}

\def\be{\begin{equation}}
\def\ee{\end{equation}}
\def\bea{\begin{eqnarray}}
\def\eea{\end{eqnarray}}

\def\s{\hat{s}}
\def\u{\hat{u}}
\def\z{v \cdot \hat{q}}

\def\ml{\hat{m}_l}
\def\ms{\hat{m}_s}

\def\mc{\hat{m}_c}

\def\lo{\hat{\lambda}_1}
\def\lt{\hat{\lambda}_2}

\def\q{\hat{q}}
\def\bxsll{$B \rightarrow X_s \ell^+ \ell^- $}
\def\absvcb{\left| V_{cb} \right|}

\catcode`@=11

\def\@citex[#1]#2{\if@filesw\immediate\write\@auxout{\string\citation{#2}}\fi
  \def\@citea{}\@cite{\@for\@citeb:=#2\do
    {\@citea\def\@citea{,\penalty\@m}\@ifundefined
      {b@\@citeb}{{\bf ?}\@warning
       {Citation `\@citeb' on page \thepage \space undefined}}%
\hbox{\csname b@\@citeb\endcsname}}}{#1}}

\def\citer{\@ifnextchar [{\@tempswatrue\@citexr}{\@tempswafalse\@citexr[]}}

\def\@citexr[#1]#2{\if@filesw\immediate\write\@auxout{\string\citation{#2}}\fi
  \def\@citea{}\@cite{\@for\@citeb:=#2\do
    {\@citea\def\@citea{--\penalty\@m}\@ifundefined
       {b@\@citeb}{{\bf ?}\@warning
       {Citation `\@citeb' on page \thepage \space undefined}}%
\hbox{\csname b@\@citeb\endcsname}}}{#1}}

\begin{document}
\twocolumn[
\begin{flushright}
DESY 98-127\\
September 1998\\
\end{flushright}
\vspace*{1.5cm}
\begin{center}
{\Large \bf
\centerline{PERTURBATIVE QCD- AND POWER-CORRECTED HADRON}
 \vspace*{0.2cm}
\centerline{SPECTRA AND SPECTRAL MOMENTS IN THE DECAY $B \to X_s \ell^+
\ell^-$}}
 \vspace*{1.5cm}
 {\large A.~Ali, G. Hiller}
\vskip0.2cm
 Deutsches Elektronen-Synchrotron DESY, Hamburg \\
Notkestra\ss e 85, D-22603 Hamburg, FRG\\

\vspace*{8.0cm}
{\large
Invited Talk; To be published in the Proceedings of the
XXIX International Conference\\ on High Energy Physics,
Vancouver, B.C., Canada,  July 23 - 29, 1998}

\end{center}]

\newpage
\title{PERTURBATIVE QCD- AND POWER-CORRECTED HADRON SPECTRA AND SPECTRAL 
MOMENTS IN THE DECAY $B \to X_s \ell^+ \ell^-$}

\author{A. ALI, G. HILLER}

\address{Deutsches Elektronen-Synchrotron DESY, 
Hamburg, Germany\\E-mail: ali@x4u2.desy.de,  ~ghiller@x4u2.desy.de}   


\twocolumn[\maketitle\abstracts{
Leading order (in $\alpha_s$) perturbative QCD and power
($1/m_b^2)$ corrections
to the hadronic invariant mass and hadron energy spectra in the decay $B
\to X_s \ell^+ \ell^-$ are reviewed in the standard model using the heavy 
quark expansion technique (HQET). In particular, the
first two hadronic moments $\langle S_H^n\rangle$ and $\langle
E_H^n\rangle$, $n=1,2$, are presented working out their sensitivity on
the HQET parameters $\lambda_1$ and $\bar{\Lambda}$. Data from the 
forthcoming B facilities can be used to measure the short-distance 
contribution in $B \to X_s \ell^+ \ell^-$ and determine the HQET 
parameters from the moments $\langle S_H^n\rangle$.
This could be combined with complementary constraints from the decay $B 
\to X \ell \nu_\ell$ to determine these parameters precisely.}]

\section{Introduction}
  The semileptonic inclusive decays $B \to X \ell^+ \ell^-$ , where
$\ell^\pm = e^\pm,\mu^\pm,\tau^\pm$ and $X$ represents a system of light 
hadronic states, offer, together with the
radiative electromagnetic penguin decays $B \to X + \gamma$, presently the
most popular testing grounds for the standard model (SM) in the flavour
sector.
%
In this contribution, we summarize the main steps in the 
derivation of the hadron spectra and hadron
spectral moments in \bxsll using perturbative QCD and the heavy quark 
expansion technique HQET 
\cite{georgi,MW,FLSold}, published recently by us \cite{AH98-1,AH98-2}.
This work, which incorporates the leading order (in $\alpha_s$) 
perturbative QCD and power ($1/m_b^2)$ corrections to the hadronic 
spectra, complements the derivation of 
the dilepton invariant mass  spectrum and the forward-backward asymmetry 
of the charged lepton \cite{AMM91}, calculated in the HQET framework 
some time ago by us in collaboration with T. Morozumi and L. 
Handoko \cite{AHHM97}. (See, also Buchalla et al. 
\cite{buchallaisidorirey}.) 
Both the hadron and dilepton spectra are needed to 
distinguish the signal ($B \to X_s \ell^+ \ell^-$) from the background 
processes and in estimating the effects of the experimental selection 
criterion. We shall concentrate here on the short-distance contribution
which can be extracted from data with the help of judicious cuts, such as 
those employed recently by the CLEO collaboration \cite{cleobsll97}.
The residual effects from the resonant (long-distance) contributions 
have been studied in these distributions elsewhere \cite{AHHM97,AH98-3}, 
to which we refer for details and references to the earlier work.

 We also underline 
the theoretical interest in measuring the first few hadronic spectral moments  
$\langle S_H^n\rangle$ and $\langle E_H^n\rangle$  
($n=1,2$). The former are sensitive to the HQET parameters
$\bar{\Lambda}$ and $\lambda_1$; we work out this dependence
numerically and argue that 
a combined analysis of the moments and spectra in \bxsll and $B \to X 
\ell \nu_\ell$ will allow to determine the HQET parameters with a high 
precision. Since these parameters are endemic to a large class of phenomena
in B decays, their precise knowledge is of great advantage in
reducing the theoretical errors in the determination of the CKM matrix 
elements $V_{td}$, $V_{ts}$, $V_{cb}$ and $V_{ub}$. 

\section{Kinematics and HQET Relations}

We start with the definition of the kinematics of the decay at the parton 
level,
$b (p_b) \to s (p_s) (+g (p_g))+\ell^{+} (p_{+})+\ell^{-}(p_{-}) $,
where $g$ denotes a gluon from the $O(\alpha_s)$ correction. 
The corresponding kinematics at the hadron level can be written as:
$B (p_B) \to X_s (p_H)+\ell^+ (p_{+})+\ell^{-} (p_{-})$.
We define by $q$ the momentum transfer to the lepton pair 
$q = p_{+}+p_{-}$ and $s \equiv q^2$ is the invariant dilepton mass squared.
 We shall also need 
the variable $u$ defined as $u \equiv -(p_b-p_+)^2+(p_b-p_{-})^2$.
The hadronic invariant mass and the hadron energy in the final state is 
denoted by  $S_H$ and $E_H$, respectively;
corresponding quantities at parton level are the 
invariant mass $s_0$ and the scaled parton energy $x_0\equiv 
\frac{E_0}{m_b}$.
{}From energy-momentum conservation, the following equalities hold in the 
$b$-quark, equivalently $B$-meson, rest frame ($v=(1,0,0,0)$):
\begin{eqnarray}
\label{eq:kin}
x_0  &=& 1- v \cdot \q \, ,
~~\s_0 = 1 -2 v \cdot \q + \s \, ,\nonumber \\
E_H &=& m_B-v \cdot q \, ,
~~S_H = m_B^2 -2 m_B v \cdot q  + s \, \, ~,
\end{eqnarray}
where dimensionless variables with a hat
are scaled by
the $b$-quark mass, e.g., $\s= \frac{s}{m_b^2}$, $\ms=\frac{m_s}{m_b}$ etc.
Here, the 4-vector $v$ denotes the velocity of both the
$b$-quark and the $B$-meson, $p_b=m_b v$ and $p_B=m_B v$.

The relation between the $B$-meson and $b$-quark mass is given by
the HQET mass relation
$m_{B}=m_b +\bar{\Lambda}-1/2 m_b (\lambda_1+3 \lambda_2)+\dots$,
where the ellipses denote terms higher order in
$1/m_b$. 
The quantity $\lambda_2$ is known precisely from the $B^* - B$
mass difference, with $\lambda_2 \simeq 0.12$ GeV$^2$. 
The other two parameters are considerably uncertain 
at present \cite{gremm,Neubert97} and are of interest here.

 The hadronic variables $E_H$ and $S_H$ can be expressed in terms of the 
partonic variables $x_0$ and $\hat{s}_0$ by the\\ following relations
\begin{eqnarray}
  E_H&=&\bar\Lambda-{\lambda_1+3\lambda_2\over2m_B}+\left(m_B-\bar
  \Lambda+
  {\lambda_1+3\lambda_2\over2m_B}\right) x_0+\dots\,, \nonumber\\
  S_H&=&m_s^2+\bar\Lambda^2+(m_B^2-2\bar\Lambda m_B+\bar\Lambda^2
  +\lambda_1+3\lambda_2)\,(\hat s_0-\hat m_s^2) \nonumber \\
  &&\qquad\qquad\mbox{}+(2\bar\Lambda
m_B-2 \bar\Lambda^2-\lambda_1-3\lambda_2) x_0
  +\dots\,~.
\nonumber
\end{eqnarray}
\noindent
The dominant non-perturbative effect on the hadron spectra is essentially
determined by the binding energy   
$\bar{\Lambda}=m_B-m_b+...$, in terms of which one has the following 
transformation:
\bea
E_0 \to E_H &=& \bar{\Lambda} + E_0 +\dots\,, \nonumber\\
s_0 \to S_H &=& s_0 + 2\bar{\Lambda} E_0 + \bar{\Lambda}^2 +\dots ~.
\eea
Thus, changing the variables of integration 
$(s_0,E_0) \to (s_H,E_0)$ and integrating over $E_0$ in the range
$\sqrt{S_H} -\bar{\Lambda} < E_0 < 1/2 m_B(S_H -2 \bar{\Lambda}
m_B^2 + m_B^2) $, one gets an invariant hadron mass spectrum 
$d \Gamma/dS_H$ in the kinematic range $ \bar{\Lambda}^2 < S_H < m_B^2$. In 
particular, already for the partonic decay $b \to s \ell^+ \ell^-$ with 
$m_s=0$, and hence $s_0=0$, one gets a non-trivial distribution in $S_H$ 
for $ \bar{\Lambda}^2 < S_H < \bar{\Lambda} m_B$.
The kinematic boundary of the distribution $d \Gamma/dS_H$ is extended by 
the bremsstrahlung process $b \to s+ g + \ell^+ \ell^-$, where now   
 $ \bar{\Lambda}m_B < S_H < m_B^2$ (with $m_s=0)$. The ${\cal O}(\alpha_s)$
contribution leads to a double logarithmic (but integrable) singularity at
$S_H=\bar{\Lambda}m_B$. 
Perturbation theory is valid for $\Delta^2 < S_H < m_B^2$, with
$\Delta^2 > \bar{\Lambda}m_B$.

\section{Matrix Element for $B \to X_s \ell^+ \ell^-$ in the Effective 
Hamiltonian Approach}
 The effective 
Hamiltonian governing the decay \bxsll is given as \cite{AHHM97}:
\begin{eqnarray}\label{heffbsll}
&&{\cal H}_{eff}(b \to s) 
= - \frac{4 G_F}{\sqrt{2}} V_{ts}^* V_{tb} \\
&&\left[ \sum_{i=1}^{6} C_i (\mu)  O_i 
+ C_7 (\mu) \frac{e}{16 \pi^2}
          \bar{s}_{\alpha} \sigma_{\mu \nu} (m_b R + m_s L) b_{\alpha}
                F^{\mu \nu} 
\right. \nonumber \\
&&\left.
+ C_9 (\mu) \frac{e^2}{16 \pi^2}\bar{s}_\alpha \gamma^{\mu} L b_\alpha
\bar{\ell} \gamma_{\mu} \ell 
+ C_{10}  \frac{e^2}{16 \pi^2} \bar{s}_\alpha \gamma^{\mu} L
b_\alpha \bar{\ell} \gamma_{\mu}\gamma_5 \ell \right] \, , \nonumber
\end{eqnarray}
where $G_F$ is the Fermi coupling constant, $L(R)=1/2(1\mp \gamma_5)$,
and $C_i$ are the Wilson coefficients.
Note that the chromo-magnetic operator does not contribute to the 
decay \bxsll in the approximation which we use here. 

The matrix element for the decay \bxsll can be factorized
into a leptonic and a hadronic part as 
\begin{equation}
        {\cal M (\mbox{\bxsll})} =
        \frac{G_F \alpha}{\sqrt{2} \pi} \, V_{ts}^\ast V_{tb} \,
        \left( {\Gamma^L}_\mu \, {L^L}^\mu
        +  {\Gamma^R}_\mu \, {L^R}^\mu \right) \, ,
\end{equation}  
with
\begin{eqnarray}
        {L^{L/R}}_\mu & \equiv &
                \bar{\ell} \, \gamma_\mu \, L(R) \, \ell \, , \\
        {\Gamma^{L/R}}_\mu & \equiv &
                \bar{s} \left[   
                R \, \gamma_\mu
                        \left( C_9^{\mbox{eff}}(\s) \mp C_{10}
                          + 2 C_7^{\mbox{eff}} \,
                        \frac{\hat{\not{q}}}{\s} \right)
                     \right. \nonumber\\
                &&\left. + 2 \hat{m}_s \, C_7^{\mbox{eff}} \, \gamma_\mu \,
                        \frac{\hat{\not{q}}}{\s} L
                \right] b \, ,
        \label{eqn:gammai}
\end{eqnarray}
with $C_7^{\mbox{eff}} \equiv C_7 - C_5/3 -C_6$.
The effective Wilson coefficient $C_9^{\mbox{eff}}(\s)$ receives
contributions from various pieces. The
resonant $c\bar{c}$ states also contribute to $C_9^{\mbox{eff}}(\s)$; hence
the contribution given below is just the perturbative part:
\begin{eqnarray}
C_9^{\mbox{eff}}(\s)=C_9 \eta(\s) + Y(\s) \, .
\end{eqnarray}
Here $\eta(\s)$ and $Y(\s)$ represent, respectively, the 
${\cal{O}}(\alpha_s)$ correction \cite{jezkuhn}
and the one loop matrix element of the
Four-Fermi operators \cite{burasmuenz,misiakE}.

With the help of the above expressions, the differential
decay width becomes on using $p_{\pm}=(E_{\pm}, \mbox{\boldmath $p_{\pm}$})$,
\begin{eqnarray}
        {\rm d} \Gamma &=& \frac{1}{2 m_B}
                \frac{{G_F}^2 \, \alpha^2}{2 \pi^2}
                \left| V_{ts}^\ast V_{tb} \right|^2
                \frac{{\rm d}^3 \mbox{\boldmath $p_+$}}{(2 \pi)^3 2 E_+}  
                \frac{{\rm d}^3 \mbox{\boldmath $p_-$}}{(2 \pi)^3 2 E_-}
                \nonumber\\
                &&\times \left( {W^L}_{\mu \nu} \, {L^L}^{\mu \nu}
                +  {W^R}_{\mu \nu} \, {L^R}^{\mu \nu} \right) \, ,
\end{eqnarray}
where $W_{\mu \nu}^{L,R}$ and $L_{\mu \nu}^{L,R}$ are the
hadronic and leptonic tensors, respectively, and can be seen in the 
literature \cite{AHHM97}. The hadronic tensor $W_{\mu\nu}^{L/R}$
is related to the discontinuity in the forward scattering amplitude,
denoted by
$T_{\mu \nu}^{L/R}$, through the relation $W^{L/R}_{\mu \nu} = 2 \, {\rm 
Im} \, T^{L/R}_{\mu \nu}$.  Transforming the integration variables
 to $\hat{s}$, $\hat{u}$ and $v \cdot \hat{q}$, one can express the
triple differential distribution in \bxsll as:
\begin{eqnarray}
      &&  \frac{{\rm d} \Gamma}{{\rm d}\u \, {\rm d}\s \, {\rm d}(\z)} = 
                \frac{1}{2 \, m_B}
                \frac{{G_F}^2 \, \alpha^2}{2 \, \pi^2} 
                \frac{{m_b}^4}{256 \, \pi^4}
                \left| V_{ts}^\ast V_{tb} \right|^2 \nonumber\\ 
              &&  \times \, 2 \, {\rm Im} 
                \left( {T^L}_{\mu \nu} \, {L^L}^{\mu \nu}
                +  {T^R}_{\mu \nu} \, {L^R}^{\mu \nu} \right) \, .
        \label{eqn:dgds}
\end{eqnarray}
Using Lorentz decomposition, the tensor $T_{\mu \nu}$ can be expanded in 
terms of three structure functions $T_i$,
\begin{equation}
        T_{\mu \nu}^{L/R} = -T_1^{L/R} \, g_{\mu \nu} + T_2^{L/R} \, 
v_\mu \, v_\nu 
                + T_3^{L/R} \, i \epsilon_{\mu \nu \alpha \beta} \, 
                        v^\alpha \, \hat{q}^\beta \, ,
\label{eq:hadrontensor}
\end{equation}
where the ones which do not contribute to the
amplitude for massless leptons have been neglected.

\section{Hadron Spectra in $B \to X_s \ell^+ \ell^-$}
We discuss first the perturbative
$O(\alpha_s)$ corrections recalling that
only the matrix element of the operator $O_9 \equiv e^2/(16
\pi^2)\bar{s}_\alpha \gamma^{\mu} L b_\alpha
\bar{\ell} \gamma_{\mu} \ell$ is subject to such corrections. The
corrected hadron energy spectrum in \bxsll can be
obtained by using the existing results in the literature
on the decay $B \to X \ell \nu_\ell$ by decomposing
the vector current in $O_9$ as $V=(V-A)/2 + (V+A)/2$.
The $(V-A)$ and $(V+A)$ currents yield
the same  hadron energy spectrum \cite{aliold}
and there is no interference term  present in this spectrum for massless
leptons. So, the correction for the vector current case
can be taken from the corresponding result for the charged $(V-A)$ case
\cite{jezkuhn}.

\begin{figure}
\center
\psfig{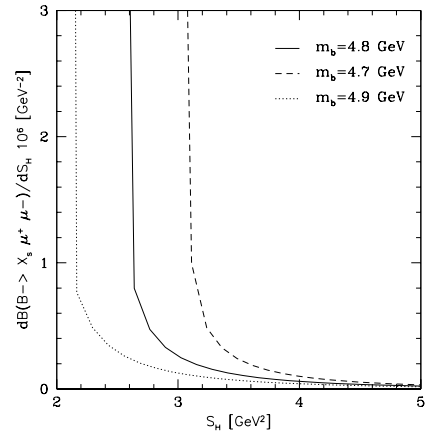} 
\caption{ The differential branching ratio $d{\cal B}(B \to X_s \ell^+
\ell^-)/dS_H$ in the hadronic invariant mass, $S_H$, shown for three values
of $m_b$ in the range where only bremsstrahlung diagrams contribute.  }
\label{fig:2}
\end{figure}

The ${\cal O}(\alpha_s)$  perturbative QCD correction
for the hadronic invariant mass is discussed next. 
As already mentioned, the decay $b \to s + \ell^+ + 
\ell^-$ yields a delta function at $\s_0 =\ms^2$ and hence only 
the bremsstrahlung diagrams $b \to s + g + \ell^+ + \ell^-$ contribute in 
the 
range $\ms^2 < \s_0 \leq 1 $. The resulting distribution $d{\cal B}(B \to 
X_s \ell^+ \ell^-)/ds_0$ in the parton model in the ${\cal O}(\alpha_s)$
approximation and the Sudakov exponentiated form 
can be seen in our paper \cite{AH98-2}. We remark that 
the Sudakov exponentiated double differential 
distribution for 
the decay $B \to X_u \ell \nu_\ell$  has been derived by Greub and 
Rey \cite{greubrey}, which we have checked and used after changing the 
normalization for 
\bxsll . The hadronic invariant mass spectrum $d{\cal B}(B \to X_s \ell^+
\ell^-)/dS_H$, shown in Fig.~\ref{fig:2}
depends rather sensitively on $m_b$ (or equivalently 
$\bar{\Lambda}$). An analogous analysis for the decay
$B \to X_u \ell \nu_\ell$ has been performed earlier, with very
similar qualitative results \cite{FLW}.

Next, we discuss the power corrections to the hadronic spectra.
The structure functions $T^{L/R}_i$ in the hadronic tensor in 
Eq.~(\ref{eq:hadrontensor})
 can be expanded in 
inverse powers of $m_b$ with the help of the 
HQET techniques \cite{georgi,MW,FLSold}.
The leading term in this expansion, i.e., ${\cal O}(m_b^0)$, reproduces the 
parton model result \cite{burasmuenz,misiakE}. In HQET, the next to leading 
power 
corrections are parameterized in terms of $\lambda_1$ and $\lambda_2$. 
After contracting the hadronic and leptonic tensors and
with the help of the kinematic identities given in Eq.~(\ref{eq:kin}),
we can make the dependence on $x_0$ and $\s_0$ explicit,
\begin{eqnarray}
       && {T^{L/R}}_{\mu \nu} \, {L^{L/R}}^{\mu \nu} = 
                {m_b}^2 \left\{ 2  (1-2 x_0+\s_0)  {T_1}^{L/R}
\right. \nonumber\\ 
&& \left.  + \left[ x_0^2 - \frac{1}{4} \u^2 - \s_0 \right] 
{T_2}^{L/R} 
                \mp  (1-2 x_0+\s_0) \u \, {T_3}^{L/R} \right\}. 
        \nonumber\\
\end{eqnarray}
 By integrating Eq.~(\ref{eqn:dgds}) over $\u$, 
the double differential power corrected spectrum can be expressed as 
\cite{AH98-2}:
 \begin{eqnarray}
\frac{{\rm d}^2 {\cal{B}}}{{\rm d} x_0 \, {\rm d}\s_0} &=& -\frac{8}{\pi} 
{\cal B}_{0}
{\mbox{Im}}\sqrt{x_0^2-\s_0} 
\left\{ (1-2 x_0+\s_0)T_1(\s_0,x_0) \right. \nonumber\\
&& \left. +\frac{x_0^2-\s_0}{3}T_2(\s_0,x_0) 
\right\} + {\cal{O}}(\lambda_i \alpha_s)
\label{doublediff} \, .
\end{eqnarray}
The structure function $T_3$ does not contribute to
the double differential distribution and we do not consider it any further. 
The functions $T_1(\s_0,x_0)$ and $T_2(\s_0,x_0)$, together with 
other details of the calculations, have been given by us elsewhere  
\cite{AH98-2}.

\begin{figure}
\center
\psfig{figure=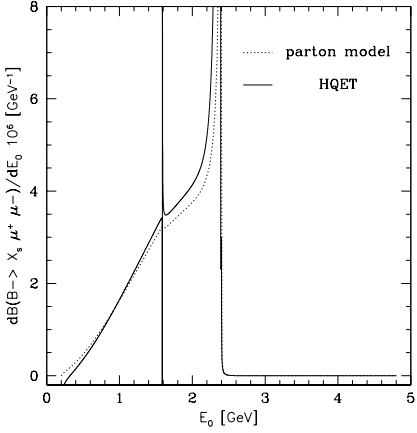,height=2.5in}
\caption{Hadron energy spectrum $d{\cal B}(B \to X_s \ell^+
\ell^-)/dE_0$ in the parton model (dotted line) and including leading
power corrections (solid line). For $m_b/2 < E_0 < m_b$ the two
distributions coincide.}
\label{fig:3}
\end{figure}

The branching ratio for \bxsll is usually expressed in terms
of the measured semileptonic branching ratio ${\cal B}_{sl}$
for the decay $B \to X_c \ell \nu_\ell$. This fixes
the normalization constant ${\cal B}_0$ to be,
\begin{equation}
        {\cal B}_0 \equiv
                {\cal B}_{sl} \frac{3 \, \alpha^2}{16 \pi^2} \frac{
    {\vert V_{ts}^* V_{tb}\vert}^2}{\absvcb^2} \frac{1}{f(\mc) \kappa(\mc)}
                \; ,
\label{eqn:seminorm}
\end{equation}
where $f(\mc)$
is the phase space factor for $\Gamma (B \rightarrow X_c \ell \nu_{\ell})$
and $\kappa(\mc)$ accounts for both the $O(\alpha_s)$ QCD 
correction to 
the semileptonic decay  width \cite{CM78} and the leading order
$(1/m_b)^2$ power correction \cite{georgi}.
The hadron energy spectrum can now be obtained by integrating over $\s_0$ with
the kinematic boundaries:
$max(\ms^2,-1+2 x_0 +4 \ml^2) \leq \s_0 \leq x_0^2$, 
$\ms \leq   x_0   \leq \frac{1}{2} (1+\ms^2-4 \ml^2)$.
The hadron energy spectrum $d{\cal B}(B \to X_s \ell^+
\ell^-)/dE_0$ in the parton model (dotted line) and including leading
power corrections (solid line) are shown in Fig.~\ref{fig:3}. For $m_b/2 < 
E_0 < m_b$ the two distributions coincide. Note that the 
$1/m_b^2$-expansion breaks down near the lower end-point of the hadron 
energy spectrum and at the $c\bar{c}$ threshold. Hence, only suitably
averaged spectra are useful for comparison with experiments in these 
regions. Apart from these regions, the
HQET and parton model spectra are remarkably close to each other. 

\section{Hadron Spectral Moments in $B \to X_s \ell^+ \ell^-$}
The lowest spectral moments
in the decay \bxsll  at the parton 
level are worked out by taking into account the
two types of corrections discussed earlier, namely the leading power $1/m_b$ 
and the perturbative ${\cal{O}}(\alpha_s)$ corrections.
To that end, we define the moments for integers $n$ and $m$:
\begin{equation}
{\cal M}^{(n,m)}_{l^{+} l^{-}} \equiv 
   {1\over {\cal B}_0}\int (\hat s_0-\hat m_s^2)^n  x_0^m\,
   {{\rm d} {\cal B}\over{\rm d}\hat s_0{\rm d} x_0}
   \,{\rm d}\hat s_0{\rm d} x_0\,,
\end{equation}
which obey 
   $\langle x_0^m(\hat s_0-\hat m_s^2)^n\rangle =
{{\cal B}_0\over {\cal B}}\,
   {\cal M}^{(n,m)}_{l^{+} l^{-}}$. These moments  
can be expanded as a double Taylor series in $\alpha_s$ and $1/m_b$:
 \begin{eqnarray}
 {\cal M}^{(n,m)}_{l^{+} l^{-}} &=&D_0^{(n,m)}+
\frac{\alpha_s}{\pi} {C_9}^2 A^{(n,m)} \nonumber\\
&& +\lo D_1^{(n,m)} + \lt D_2^{(n,m)} \,\, ,
\end{eqnarray}
with a further decomposition of $D_i^{(n,m)}$, $i=0,1,2$, into pieces from 
different Wilson coefficients:
\begin{eqnarray}
\label{momentexp}
D_i^{(n,m)}=\alpha_i^{(n,m)} {C_7^{{\mbox{eff}}}}^2+
\beta_i^{(n,m)} C_{10}^2+
\gamma_i^{(n,m)} C_7^{{\mbox{eff}}} +\delta_i^{(n,m)}. \nonumber\\
\end{eqnarray}
The terms $\gamma_i^{(n,m)}$ and $\delta_i^{(n,m)}$ in              
Eq.~(\ref{momentexp}) result from the terms proportional 
to ${\it{Re}}(C_9^{{\mbox{eff}}})C_7^{{\mbox{eff}}}$ and
$|C_9^{{\mbox{eff}}}|^2$  in  Eq.~(\ref{doublediff}), respectively.   
The explicit expressions for   
$\alpha_i^{(n,m)},\beta_i^{(n,m)},  \gamma_i^{(n,m)}, \delta_i^{(n,m)}$
are given in our paper \cite{AH98-2}.

The leading perturbative contributions for the hadronic invariant mass and 
hadron energy 
moments can be obtained analytically,  
\begin{eqnarray}
A^{(0,0)}&=&\frac{25-4 \pi^2}{9} \, ,
~~A^{(1,0)}=\frac{91}{675} \, ,
~~A^{(2,0)}=\frac{5}{486} \, ,\nonumber \\
A^{(0,1)}&=&\frac{1381-210 \pi^2}{1350} \, ,
A^{(0,2)}=\frac{2257-320 \pi^2}{5400} \, .
\label{eq:A10}
\end{eqnarray}
The zeroth moment $n=m=0$ is needed for the normalization;  
the result for $A^{(0,0)}$ was first derived by Cabibbo and Maiani 
\cite{CM78}.
Likewise, the first mixed moment $A^{(1,1)}$ can be extracted from 
the results for the decay $B \to X \ell \nu_{\ell}$ \cite{FLS} after 
changing the normalization, $A^{(1,1)}=3/50$.
For the lowest order parton model contribution 
$D_0^{(n,m)}$, we find, in agreement 
with \cite{FLS}, that the first two hadronic invariant mass moments 
$\langle \s_0-\ms^2 \rangle, \, \langle(\s_0-\ms^2)^2 \rangle$ and the first 
mixed moment $\langle x_0 (\s_0-\ms^2) \rangle$ 
vanish: $D_0^{(n,0)}=0$, for $n=1,2$ and $D_0^{(1,1)}=0 $ .

Using the expressions for the HQET moments derived by us \cite{AH98-2}, 
we  present the numerical results for the hadronic moments in \bxsll.
The parameters used are : $m_s = 0.2 ~\mbox{GeV},m_c = 1.4 ~\mbox{GeV},
m_b = 4.8 ~\mbox{GeV},m_t = 175 \pm 5 ~\mbox{GeV},\mu ={m_{b}}^{+m_{b}}_{-m_{b}/2},\alpha_s (m_Z) = 0.117 \pm 0.005 ~,\alpha^{-1} = 129$.
We find for the short-distance hadronic moments, 
valid up to ${\cal{O}}(\alpha_s/m_B^2,1/m_B^3)$:
\begin{eqnarray}
\label{eq:corrs}
   \langle S_H\rangle&=&m_B^2 (\frac{m_s^2}{m_B^2}
+0.093 \frac{\alpha_s}{\pi} 
-0.069 \frac{\bar{\Lambda}}{m_B} \frac{\alpha_s}{\pi} \nonumber\\
&&+0.735 \frac{\bar{\Lambda}}{m_B}+0.243 \frac{\bar{\Lambda}^2}{m_B^2}
+ 0.273 \frac{\lambda_1}{m_B^2}-0.513\frac{\lambda_2}{m_B^2}) \nonumber 
\, ,\\ %
\label{eq:hadmoments}
 \langle S_H^2\rangle&=&m_B^4 (0.0071 \frac{\alpha_s}{\pi} 
+0.138 \frac{\bar{\Lambda}}{m_B} \frac{\alpha_s}{\pi} \nonumber\\
&&+0.587\frac{\bar{\Lambda}^2}{m_B^2}
-0.196 \frac{\lambda_1}{m_B^2}) \, ,\\
   \langle E_H\rangle&=& 0.367 m_B  (1+0.148 \frac{\alpha_s}{\pi} 
-0.352 \frac{\bar{\Lambda}}{m_B} \frac{\alpha_s}{\pi}
+1.691 \frac{\bar{\Lambda}}{m_B} \nonumber\\
&&+0.012\frac{\bar{\Lambda}^2}{m_B^2}
+ 0.024 \frac{\lambda_1}{m_B^2}+1.070\frac{\lambda_2}{m_B^2}) \, ,\nonumber \\
 \langle E_H^2\rangle&=&0.147 m_B^2 (1+0.324 \frac{\alpha_s}{\pi} 
-0.128 \frac{\bar{\Lambda}}{m_B} \frac{\alpha_s}{\pi}
+2.954 \frac{\bar{\Lambda}}{m_B} \nonumber\\
&&+2.740\frac{\bar{\Lambda}^2}{m_B^2}
-0.299 \frac{\lambda_1}{m_B^2}+0.162\frac{\lambda_2}{m_B^2}) \, ,\nonumber
\end{eqnarray}
where the numbers shown correspond to the central values of the parameters.

\begin{figure}
\center
\psfig{figure=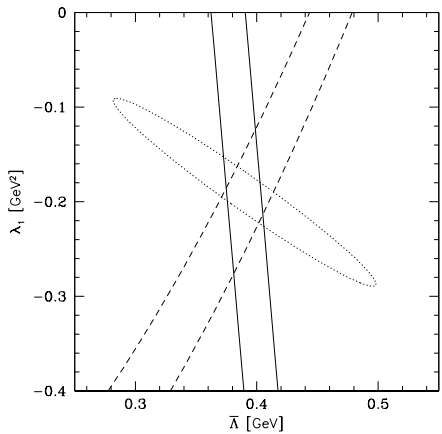,height=2.5in}
\caption{$\langle S_H \rangle$ (solid bands) and $\langle S_H^2\rangle$
(dashed bands) correlation in $(\lambda_1$-$\bar{\Lambda})$ space 
for the decay $B \to X_s \ell^+\ell^-$. The correlation from
the analysis of the decay $B \to X\ell \nu_\ell$ by Gremm et al. 
\protect\cite{gremm} is shown as an ellipse. }         
\label{fig:4}                                                          
\end{figure}

The dependence of the hadronic moments given in
Eq.~(\ref{eq:hadmoments}) on the HQET parameters
$\lambda_1$ and $\bar{\Lambda}$ has been worked out numerically. In doing 
this, the theoretical 
errors on these moments following from the errors on the input
parameters $m_t$, $\alpha_s$ and the scale $\mu$ have been estimated 
by varying these parameters in the indicated $\pm 1\sigma$ ranges, one at a 
time, and adding the individual errors in quadrature. 
The correlations on the HQET parameters $\lambda_1$ and $\bar{\Lambda}$
which follow from (assumed) fixed
values of the hadronic invariant mass moments  $\langle S_H \rangle$ 
and  $\langle S_H^2 \rangle$ (calculated using $\bar{\Lambda}=0.39 \, 
{\mbox{GeV}}$,
$\lambda_1=-0.2 \, {\mbox{GeV}}^2$ and $\lambda_2=0.12 \, {\mbox{GeV}}^2$)
are shown in Fig.~\ref{fig:4} (for the decay $B \to X_s \mu^+ \mu^-$). The
($\lambda_1$-$\bar{\Lambda})$ correlation  from the analysis of Gremm et al. 
\cite{gremm} for the electron energy spectrum in $B \to X \ell \nu_\ell$
is shown as an ellipse in this figure. With the measurements of
$\langle S_H \rangle$ and  $\langle S_H^2 \rangle$ in the decay $B \to 
X_s \ell^+ \ell^-$, one has to solve the experimental numbers on the 
l.h.s. of Eq.~(\ref{eq:corrs}) for $\lambda_1$ and $\bar{\Lambda}$. It is, 
however, clear that the constraints 
from the decays \bxsll~and $B \to X \ell \nu_\ell$ are complementry.
 Using the CLEO cuts on hadronic and dileptonic
masses \cite{cleobsll97}, we estimate that $O(200)$ \bxsll ($\ell=e,\mu$) 
events will be available per $10^7$ $B\bar{B}$ hadrons \cite{AH98-2}.
So, there will be plenty of \bxsll decays in the forthcoming B facilities
to measure the correlation shown in Fig.~\ref{fig:4}. 
 
 Of course, the utility 
of the hadronic moments calculated above is only in conjunction
with the experimental cuts which could effectively remove the resonant
(long-distance) contributions. The optimal experimental cuts in \bxsll 
remain to be 
defined, but for the cuts used by the CLEO collaboration we have studied
the effects in the HQET-like Fermi motion (FM) model \cite{aliqcd}.
We find that the hadronic moments in the HQET and FM model are very similar
and CLEO-type cuts remove the bulk of the $c \bar{c}$ resonant contributions
\cite{AH98-2}.

In summary, we have calculated the dominant contributions
to the hadron spectra and spectral moments in $B \to
X_s \ell^+ \ell^-$  including
contributions up to terms of ${\cal 
O}(\alpha_s/m_B^2,1/m_B^3)$. 
We have presented the results on the spectral hadronic moments $\langle 
E_H^n \rangle$
and $\langle S_H^n \rangle$ for $n=1,2$ and have worked out their dependence
on the HQET parameters $\bar{\Lambda}$ and $\lambda_1$.
The correlations in \bxsll are shown to be different
than the ones in the semileptonic decay $B \to X \ell \nu_\ell$. This
complementarity allows, in principle, a powerful method to determine them 
precisely from data on $B \to X \ell \nu_\ell$ and \bxsll in forthcoming
high luminosity $B$ facilities.

\end{document}